\documentclass{article}
\usepackage{spconf,amsmath,graphicx,hyperref}

\usepackage{comment}
\usepackage{multirow}
\usepackage{booktabs}
\usepackage{color}
\usepackage{subcaption}

\title{
Listening without Looking:\\
Modality Bias in Audio-Visual Captioning
}
\name{
Yuchi Ishikawa$^{1,2}$,
Toranosuke Manabe$^{1,2}$,
Tatsuya Komatsu$^{1}$,
Yoshimitsu Aoki${^2}$
}
\address{
$^1$LY Corporation,
$^2$Keio University
}
\begin{document}
%
\maketitle

\begin{abstract}
Audio-visual captioning aims to generate holistic scene descriptions by jointly modeling sound and vision.
While recent methods have improved performance through sophisticated modality fusion,
it remains unclear to what extent the two modalities are complementary in current audio-visual captioning models and how robust these models are when one modality is degraded.
We address these questions by conducting systematic modality robustness tests on LAVCap,
a state-of-the-art audio-visual captioning model,
in which we selectively suppress or corrupt the audio or visual streams to quantify sensitivity and complementarity.
The analysis reveals a pronounced bias toward the audio stream in LAVCap.
To evaluate how balanced audio-visual captioning models are in their use of both modalities,
we augment AudioCaps with textual annotations that jointly describe the audio and visual streams, yielding the \textit{AudioVisualCaps} dataset.
In our experiments, we report LAVCap baseline results on AudioVisualCaps.
We also evaluate the model under modality robustness tests on AudioVisualCaps
and the results indicate that LAVCap trained on AudioVisualCaps exhibits less modality bias
than when trained on AudioCaps.
\end{abstract}

\section{Introduction}

Automatic captioning has emerged as a core research problem at the intersection of computer vision,
audio processing, and natural language processing.
Advances in multimodal learning and the availability of large‑scale datasets have broadened the scope
beyond image/video-only or audio-only captioning~\cite{vinyals2015show,
venugopalan2015sequence,drossos2020clotho}
to \emph{audio-visual captioning} (AVC)~\cite{kim2024avcap,liu2022visually,xu2025mitigating}, which jointly exploits both audio and vision
\footnote{%
Although some studies~\cite{liu2022visually,xu2025mitigating} distinguish \textit{audio-visual captioning} from \textit{visually aware audio captioning} or \textit{visual-guide audio captioning},
where visual signals serve as cues for audio captioning,
we intentionally adopt the term ``audio-visual captioning''
to emphasize that both modalities are treated equally in generating captions.
}.
Compared with single‑modality captioning,
AVC aims to produce more holistic scene descriptions that better reflect human perception.
Its ability to integrate auditory and visual cues also makes it useful for downstream tasks such as multimodal retrieval and audiovisual scene understanding.

Recent studies~\cite{rho2025lavcap,kim2024avcap,xu2025mitigating} have primarily investigated how to fuse audio and visual information effectively.
For instance, AVCap~\cite{kim2024avcap} uses CAV-MAE~\cite{gong2022contrastive} to extract joint embeddings,
which are then fed as tokens into a text encoder.
LAVCap~\cite{rho2025lavcap}, a large language model (LLM)-based AVC framework,
employs an optimal-transport attention module to integrate audio and visual features.
These fusion strategies yield richer and more detailed captions than unimodal approaches.

However, two research questions remain underexplored:
\textbf{RQ1 (Complementarity):} To what extent do current AVC models rely on audio versus visual signals, and are these signals used in a complementary manner?
\textbf{RQ2 (Robustness):} How robust are AVC models to degradation or absence of one modality, and can one modality compensate for the other?

To address these questions,
we evaluate a state‑of‑the‑art AVC model
under modality robustness tests.
We examine how performance changes when we suppress or corrupt either the audio or the visual stream. 
The results reveal an asymmetric reliance on the two modalities.
The model remains relatively robust to visual perturbations but degrades substantially under audio perturbations. 
We conclude that the examined model exhibits a strong bias toward the auditory modality.
We hypothesize that this bias stems in part from commonly used AVC datasets (e.g., AudioCaps~\cite{kim2019audiocaps}) that place comparatively greater emphasis on the audio stream.

\begin{table*}[t]
\small
\centering
\caption{
\textbf{Audio–visual captioning results on AudioCaps under the modality robustness test.}
We evaluate LAVCap as the audio–visual captioning model.
Visual inputs: original frames, black/white (blank), random noise image and frames shuffled in the dataset.
Audio inputs: original audio clips, silence, Gaussian noise, audio clips shuffled in the dataset.
}
\vspace{-2mm}
\begin{tabular}{l l c c c c c c c}
\toprule[1.2pt]
\textbf{Audio} & \textbf{Visual} & \textbf{BLEU-1 $\uparrow$} & \textbf{BLEU-4 $\uparrow$} & \textbf{ROUGE\text{-}L $\uparrow$} & \textbf{METEOR $\uparrow$} & \textbf{CIDEr $\uparrow$} & \textbf{SPICE $\uparrow$} & \textbf{SPIDEr $\uparrow$} \\
\midrule[0.5pt]
Original & Original & 0.721 & 0.283 & 0.510 & 0.256 & 0.805 & 0.182 & 0.494 \\
\midrule[0.5pt]
Original & Black     & 0.672 & 0.248 & 0.483 & 0.234 & 0.685 & 0.161 & 0.423 \\
Original & White     & 0.672 & 0.245 & 0.482 & 0.236 & 0.674 & 0.164 & 0.419 \\
Original & Noise image & 0.675 & 0.251 & 0.489 & 0.238 & 0.697 & 0.166 & 0.432 \\
Silent   & Original  & 0.537 & 0.145 & 0.373 & 0.164 & 0.333 & 0.093 & 0.213 \\
Gaussian & Original  & 0.319 & 0.091 & 0.223 & 0.092 & 0.226 & 0.057 & 0.142 \\
\midrule[0.5pt]
Original & Shuffled Visual & 0.623 & 0.210 & 0.452 & 0.220 & 0.566 & 0.152 & 0.359 \\
Shuffled Audio & Original & 0.408 & 0.076 & 0.282 & 0.109 & 0.140 & 0.049 & 0.094 \\
\bottomrule[1.2pt]
\end{tabular}
\vspace{-2mm}
\label{tab:preliminary}
\end{table*}
\begin{figure*}[t]
    \centering
    \small
    \includegraphics[width=0.95\linewidth]{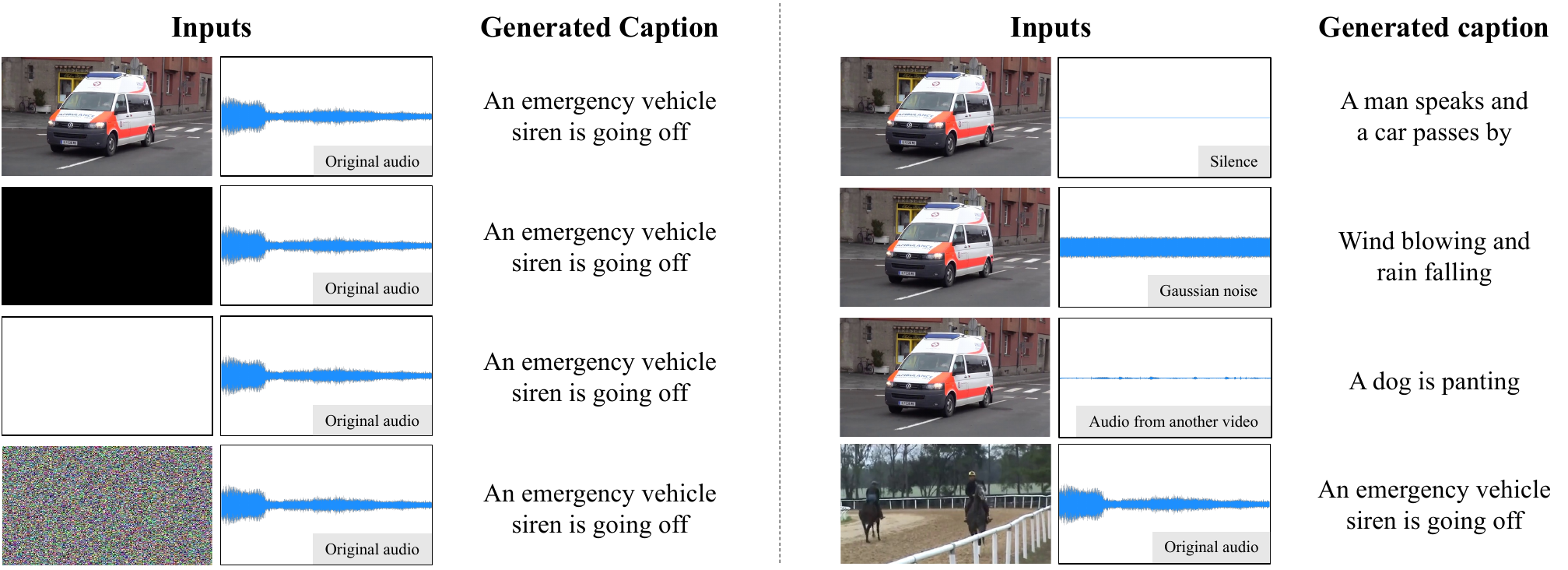}
    \vspace{-1mm}
    \caption{
    \textbf{Caption examples under modality robustness tests.}
    }
    \vspace{-2mm}
    \label{fig:preliminary}

\end{figure*}

Furthermore, to evaluate how balanced AVC models are in their use of both modalities,
we augment the AudioCaps dataset with textual annotations that describe both the audio and visual streams.
We refer to the resulting dataset as \textit{AudioVisualCaps}\footnote{We will release the dataset upon acceptance.}.
To construct these annotations, we first obtain per‑modality captions.
In particular, we use the original captions in AudioCaps as the audio captions,
while we generate visual captions using image captioning models.
For the test set, we merge the per‑modality captions with an LLM‑based procedure and manually verify the outputs to obtain the final reference captions.
For the training set, we forgo merging due to annotation costs and instead use the two per‑modality captions as separate references.
Using this dataset, we train LAVCap on AudioVisualCaps and report baseline results.

To summarize, our main contributions are threefold:
(i) An empirical analysis of modality bias in state‑of‑the‑art audio-visual captioning under modality robustness tests;
(ii) A set of textual annotations for the AudioCaps dataset that jointly describe the audio and visual streams (\textit{AudioVisualCaps}); and
(iii) Baseline results with LAVCap trained on AudioVisualCaps, demonstrating reduced modality bias.

\section{Modality Robustness Tests}
\label{sec:pre}

\subsection{Experimental setup}

In this work, we first address the two research questions introduced in the previous section:
\textbf{RQ1 (Complementarity).} To what extent do current AVC models rely on audio versus visual signals, and are these signals used in a complementary manner?
\textbf{RQ2 (Robustness).} How robust are AVC models to degradation or absence of one modality, and can one modality compensate for the other?
To investigate these questions, we conduct modality robustness tests.

In the modality robustness tests, we use LAVCap~\cite{rho2025lavcap} trained on AudioCaps~\cite{kim2019audiocaps}
and measure how the model’s performance changes when we perturb one of the input modalities
(e.g., by suppressing it, adding noise, or swapping it with that of another sample).
We report standard captioning metrics (BLEU‑1/4~\cite{papineni2002bleu},
ROUGE‑L~\cite{lin2004rouge}, METEOR~\cite{banerjee2005meteor}, CIDEr~\cite{vedantam2015cider},
SPICE~\cite{anderson2016spice}, and SPIDEr~\cite{liu2017improved}).

For visual perturbations, we replace the input image with an all‑white/all‑black image or with a random‑noise image.
For audio perturbations, we replace the input audio with silence or Gaussian noise.
Note that because LAVCap employs an optimal‑transport attention module, it cannot accept a truly missing modality.
To emulate a missing‑modality condition, we therefore feed all‑white/all‑black images or silent audio.

Additionally, on the AudioCaps test set we evaluate a swapping protocol
in which either the audio or the visual stream of each sample is replaced with that of a different sample.
Although similar experiments have been explored in prior work~\cite{xu2025mitigating} on visually aware audio captioning,
it did not shuffle the audio modality.
In contrast, we explicitly shuffle the audio stream as well.

\begin{figure}[t]
  \centering
  \includegraphics[width=\linewidth]{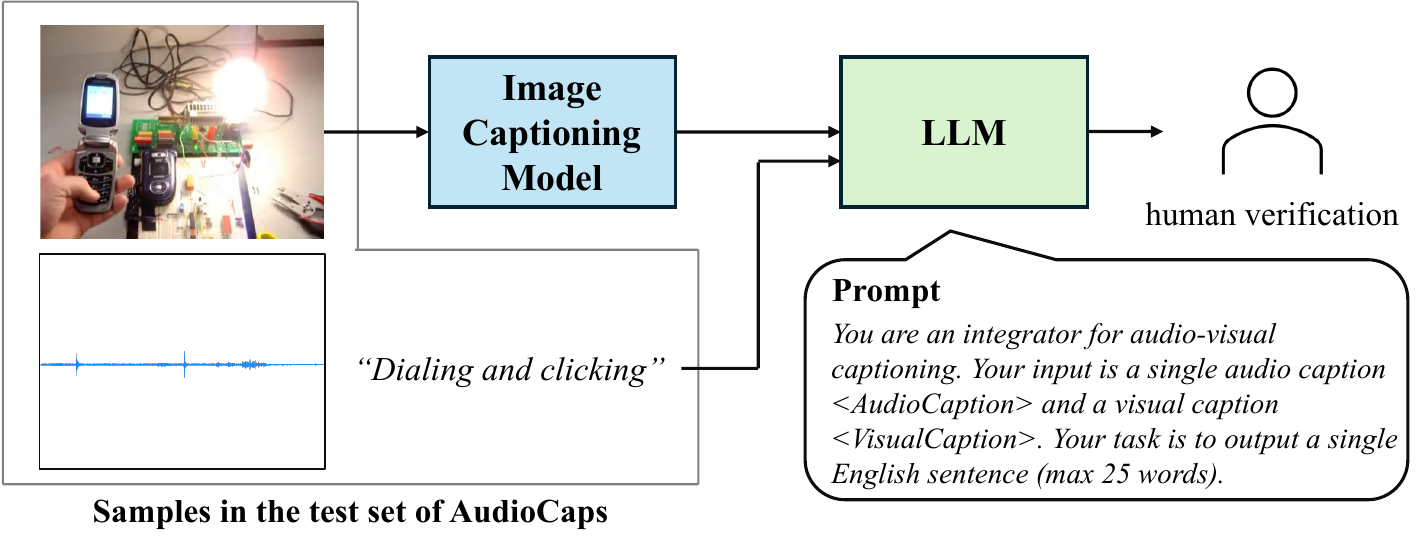}
  \caption{\textbf{Annotation pipeline for the test set.}
  We first generate visual captions from video frames.
  Next, an LLM integrates them with audio-focused captions from AudioCaps.
  Finally, we manually verify the results to produce the final captions that capture both visual and audio modalities.
  }
  \label{fig:dataset_gen}
  \vspace{-1mm}
\end{figure}

\subsection{Results}

Table~\ref{tab:preliminary} summarizes the modality robustness results.
LAVCap is markedly more sensitive to perturbations on the \emph{audio} stream than on the \emph{visual} stream.
Under visual blanking or random‑noise images, performance degrades only moderately (e.g., CIDEr drops from $0.805$ to $0.674$ and BLEU‑4 from $0.283$ to $0.245$).
By contrast, audio corruption has a much larger impact: replacing audio with silence or Gaussian noise reduces CIDEr to $0.333$ and $0.226$, respectively, with similar trends for BLEU‑4 and SPIDEr.

When we explicitly break cross‑modal alignment by shuffling one modality across the test set, the asymmetry becomes even clearer.
Shuffling the visual stream yields a moderate decline (e.g., a $30\%$ drop in CIDEr and a $27\%$ drop in SPIDEr, relative to the clean condition),
whereas shuffling the audio stream is far more severe (e.g., an $83\%$ drop in CIDEr and an $81\%$ drop in SPIDEr).

The small gaps among the blank images and random-noise images further suggest that,
when audio is available, the model exploits little fine‑grained visual content, whereas corrupting or misaligning audio severely impairs generation.

Fig.~\ref{fig:preliminary} presents qualitative examples under these modality robustness tests.
LAVCap is not affected when the visual input is missing or noisy, often producing similar captions,
but it generates entirely different captions when the audio stream is corrupted.
With missing audio, the model appears to rely on the visual stream.
However, the resulting captions still fail to capture fine‑grained visual details.
In addition, replacing the visual stream with frames from a different sample changes the output only marginally,
whereas replacing the audio stream substantially alters the resulting caption.

These results point to an asymmetric reliance on the two modalities.
The evaluated model remains relatively robust to visual perturbations but degrades substantially under audio perturbations.
Accordingly, for RQ1 (Complementarity) we find that the model relies predominantly on the audio stream, with limited complementary use of visual cues.
For RQ2 (Robustness) it is robust to visual degradation but brittle to audio degradation, and the visual stream only partially compensates for corrupted or missing audio.
We hypothesize that this bias is partly induced by AudioCaps that places greater emphasis on the audio stream.

\section{AudioVisualCaps}

\begin{table}[t]
\centering
\small

\caption{
\textbf{Comparison of caption statistics between AudioCaps and AudioVisualCaps.}
}
\vspace{-2mm}

\begin{tabular}{llll}
\toprule[1.2pt]
test split      & \#words/caption & \#vocabulary \\
\midrule[0.5pt]
AudioCaps       & 10.27           & 1319         \\
AudioVisualCaps & 17.22           & 2399         \\
\bottomrule[1.2pt]
\end{tabular}
\label{tab:statistics}
\vspace{-1mm}
\end{table}

\begin{figure}[t]
    \centering
    \small
    \includegraphics[width=0.95\linewidth]{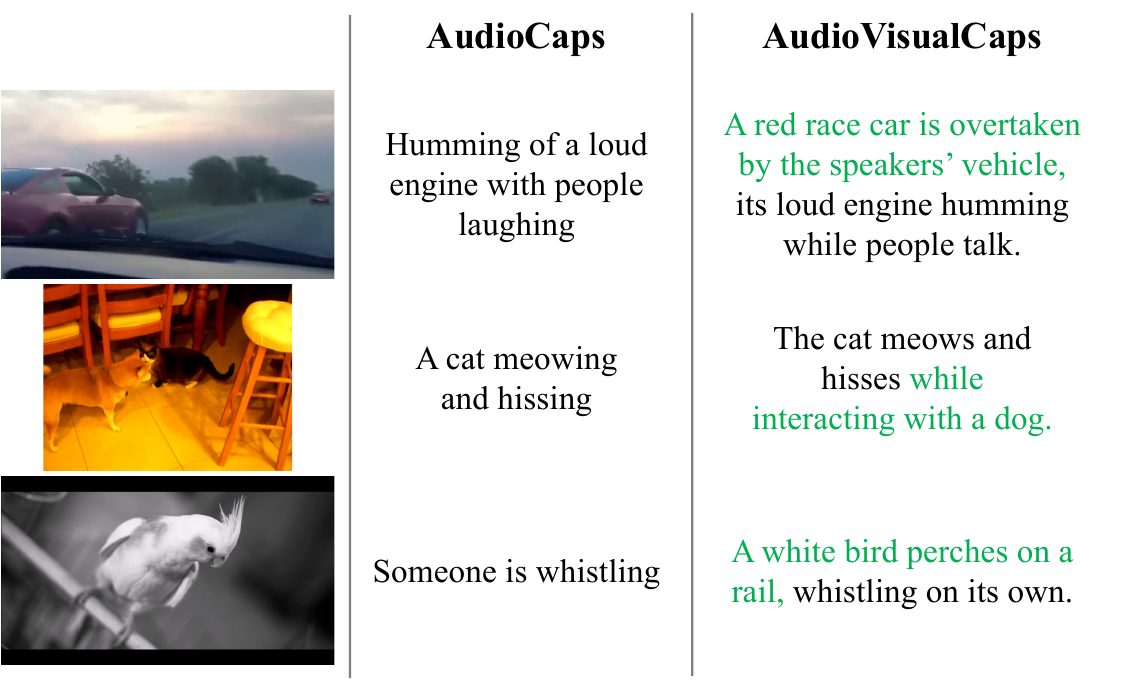}

    \vspace{-1mm}
    \caption{
    \textbf{Samples from the AudioVisualCaps test set.}
    In contrast to AudioCaps, our captions capture both audio and visual information (visual content is highlighted in green).
    }
    \label{fig:audiovisualcaps}
    \vspace{-1mm}

\end{figure}

\begin{table*}[t]
\centering
\caption{
\textbf{The baseline results of LAVCap on AudioVisualCaps.}
In the AudioVisualCaps training set,
captions are provided for each modality, and there are no merged captions unlike the test set.
Therefore, we select captions for training as following:
(random) randomly selects either the image caption or the audio caption each time,
(concat) concatenates the audio and visual captions,
(and) connects the audio and visual captions with "and".
}
\vspace{-2mm}
\begin{tabular}{lccccccc}
\toprule[1.2pt]
\textbf{Training Dataset} & \textbf{BLEU-1 $\uparrow$} & \textbf{BLEU-4 $\uparrow$} & \textbf{ROUGE\text{-}L $\uparrow$} & \textbf{METEOR $\uparrow$} & \textbf{CIDEr $\uparrow$} & \textbf{SPICE $\uparrow$} & \textbf{SPIDEr $\uparrow$} \\
\midrule[0.5pt]
AudioCaps                & 0.327 & 0.080 & 0.300 & 0.138 & 0.204 & 0.102 & 0.153 \\
AudioVisualCaps (random) & 0.448 & 0.160 & \textbf{0.376} & 0.155 & 0.392 & 0.115 & 0.254 \\
AudioVisualCaps (concat) & 0.595 & 0.174 & 0.363 & 0.211 & 0.638 & 0.159 & 0.399 \\
AudioVisualCaps (and)    & \textbf{0.599} & \textbf{0.181} & 0.368 & \textbf{0.214} & \textbf{0.665} & \textbf{0.166} & \textbf{0.416} \\
\bottomrule[1.2pt]
\end{tabular}
\vspace{-1mm}
\label{tab:existing}
\end{table*}

\begin{table*}[t]
\small
\centering
\caption{
\textbf{The performance of LAVCap on AudioVisualCaps under the modality robustness test.}
}
\vspace{-2mm}
\begin{tabular}{l l c c c c c c c}
\toprule[1.2pt]
\textbf{Audio} & \textbf{Visual} & \textbf{BLEU-1 $\uparrow$} & \textbf{BLEU-4 $\uparrow$} & \textbf{ROUGE\text{-}L $\uparrow$} & \textbf{METEOR $\uparrow$} & \textbf{CIDEr $\uparrow$} & \textbf{SPICE $\uparrow$} & \textbf{SPIDEr $\uparrow$} \\
\midrule[0.5pt]
Original & Original  & 0.599 & 0.181 & 0.368 & 0.214 & 0.665 & 0.166 & 0.416 \\
\midrule[0.5pt]
Original & Black     & 0.316 & 0.000 & 0.215 & 0.055 & 0.018 & 0.031 & 0.025 \\
Original & White     & 0.317 & 0.000 & 0.215 & 0.055 & 0.018 & 0.031 & 0.025 \\
Original & Noise Image & 0.318 & 0.008 & 0.216 & 0.056 & 0.020 & 0.031 & 0.026 \\
Silent   & Original  & 0.362 & 0.029 & 0.271 & 0.103 & 0.141 & 0.060 & 0.101 \\
Gaussian & Original  & 0.366 & 0.029 & 0.275 & 0.104 & 0.136 & 0.061 & 0.099 \\
\midrule[0.5pt]
Original & Shuffled Visual & 0.315 & 0.027 & 0.242 & 0.076 & 0.049 & 0.024 & 0.036 \\
Shuffled Audio & Original  & 0.368 & 0.030 & 0.276 & 0.105 & 0.137 & 0.058 & 0.098 \\
\bottomrule[1.2pt]
\end{tabular}
\label{tab:main}
\vspace{-1mm}
\end{table*}

From the experiments so far, we observe that AudioCaps is biased toward the audio stream,
which limits our ability to assess how well current AVC models balance the two modalities.
To address this limitation, we propose \emph{AudioVisualCaps}, an extension of the AudioCaps dataset~\cite{kim2019audiocaps}.
A key feature is that each sample is paired with a caption that jointly reflects information from both the audio and visual streams.
Creating such annotations entirely by hand would be prohibitively expensive.
Inspired by~\cite{wang2023internvid,ishikawa2025language},
we therefore introduce a semi‑automatic pipeline that leverages captioning models and an LLM to generate the captions (as illustrated in Fig.~\ref{fig:dataset_gen}).

The pipeline is as follows: 
For each AudioCaps sample, we first obtain per‑modality captions.
Herein, we use BLIP‑2~\cite{li2023blip} for image captioning.
For audio captions, we adopt the AudioCaps annotations,
which primarily describe the audio content.
Then, for the \emph{test} set, we use \texttt{gpt-oss-20b}~\cite{agarwal2025gpt},
a state-of-the-art open-weight LLM,
to merge the two per‑modality captions into a single caption that captures information from \emph{both} streams, followed by a lightweight human review to obtain the final references.
For the \emph{training} set, we forgo merging due to computational and annotation costs and instead use the two per‑modality captions as separate references.

Table~\ref{tab:statistics} shows the comparison of caption statistics between AudioCaps and our AudioVisualCaps.
We also provide
examples of the resulting annotations in Fig.~\ref{fig:audiovisualcaps}. 
As shown, the AudioVisualCaps annotations provide more detailed captions that draw on both modalities.

\section{Experiments}
In this section, we train LAVCap on the proposed AudioVisualCaps dataset and report baseline results (Sec.~\ref{sec:baseline}).
We also conduct modality robustness tests on AudioVisualCaps and compare the model's behavior with that observed on AudioCaps (Sec.~\ref{sec:modality_perturbation_on_avc}).
The setting of modality robustness tests is identical to that in Sec.~\ref{sec:pre}.
We follow the training configuration of~\cite{rho2025lavcap}, except for the number of training epochs, which we set to 25.
We use the same evaluation metrics as before.

\subsection{Performance on AudioVisualCaps}
\label{sec:baseline}

Table~\ref{tab:existing} reports LAVCap baselines on AudioVisualCaps.
We find that LAVCap trained on AudioCaps performs substantially worse on the AudioVisualCaps test set.
We hypothesize that training on AudioCaps biases the representation toward the audio modality, leading to underutilization of visual information.
By contrast, training on AudioVisualCaps yields large improvements, reflecting the model's ability to integrate both audio and visual streams. 
Furthermore, although the AudioVisualCaps training set includes neither LLM‑merged nor manually annotated captions,
we find that simply concatenating the two per‑modality captions (i.e., joining them with ``and'') already improves performance.
We therefore adopt this concatenated variant in the following experiment.

\subsection{Modality Robustness Tests on AudioVisualCaps}
\label{sec:modality_perturbation_on_avc}

Next, we evaluate LAVCap trained on AudioVisualCaps in the modality robustness tests.
Table~\ref{tab:main} reports LAVCap's performance on AudioVisualCaps.
Compared to the asymmetric results observed on AudioCaps (Sec.~\ref{sec:pre}),
the results here are symmetric (i.e. when the audio modality is missing or shuffled, performance degrades, and vice versa for the visual modality.)
This indicates that AudioVisualCaps exhibits less modality bias than AudioCaps and encourages a more balanced use of audio and visual modalities.

\section{Conclusion}

In this work, we investigated modality bias in audio-visual captioning through modality robustness tests.
The results reveal that a state-of-the-art audio-visual captioning model relies predominantly on the audio stream.
While the model remains relatively robust to visual degradation, it degrades markedly under audio corruption, indicating asymmetry in complementarity (RQ1) and robustness (RQ2).
To better assess and encourage balanced use of both modalities,
we augmented the AudioCaps dataset with captions that jointly describe the audio and visual streams, yielding the \emph{AudioVisualCaps} dataset.
Our semi-automatic annotation pipeline offers a scalable path toward constructing balanced audio-visual references while keeping human effort modest.
In our experiments, training LAVCap on AudioVisualCaps reduced modality bias compared to training on AudioCaps,
and yielded more symmetric performance under modality robustness tests.
We believe that our analyses and the AudioVisualCaps dataset will play a key role in advancing future research on audio-visual captioning.

\clearpage

\bibliographystyle{IEEEbib}
\bibliography{main}

\begin{thebibliography}{10}

\bibitem{vinyals2015show}
Oriol Vinyals, Alexander Toshev, Samy Bengio, and Dumitru Erhan,
\newblock ``Show and tell: A neural image caption generator,''
\newblock in {\em Proceedings of the IEEE conference on computer vision and pattern recognition}, 2015, pp. 3156--3164.

\bibitem{venugopalan2015sequence}
Subhashini Venugopalan, Marcus Rohrbach, Jeffrey Donahue, Raymond Mooney, Trevor Darrell, and Kate Saenko,
\newblock ``Sequence to sequence-video to text,''
\newblock in {\em Proceedings of the IEEE international conference on computer vision}, 2015, pp. 4534--4542.

\bibitem{drossos2020clotho}
Konstantinos Drossos, Samuel Lipping, and Tuomas Virtanen,
\newblock ``Clotho: An audio captioning dataset,''
\newblock in {\em ICASSP 2020-2020 IEEE International Conference on Acoustics, Speech and Signal Processing (ICASSP)}. IEEE, 2020, pp. 736--740.

\bibitem{kim2024avcap}
Jongsuk Kim, Jiwon Shin, and Junmo Kim,
\newblock ``Avcap: Leveraging audio-visual features as text tokens for captioning,''
\newblock {\em arXiv preprint arXiv:2407.07801}, 2024.

\bibitem{liu2022visually}
Xubo Liu, Qiushi Huang, Xinhao Mei, Haohe Liu, Qiuqiang Kong, Jianyuan Sun, Shengchen Li, Tom Ko, Yu~Zhang, Lilian~H Tang, et~al.,
\newblock ``Visually-aware audio captioning with adaptive audio-visual attention,''
\newblock {\em arXiv preprint arXiv:2210.16428}, 2022.

\bibitem{xu2025mitigating}
Le~Xu, Chenxing Li, Yong Ren, Yujie Chen, Yu~Gu, Ruibo Fu, Shan Yang, and Dong Yu,
\newblock ``Mitigating audiovisual mismatch in visual-guide audio captioning,''
\newblock {\em arXiv preprint arXiv:2505.22045}, 2025.

\bibitem{rho2025lavcap}
Kyeongha Rho, Hyeongkeun Lee, Valentio Iverson, and Joon~Son Chung,
\newblock ``Lavcap: Llm-based audio-visual captioning using optimal transport,''
\newblock in {\em ICASSP 2025-2025 IEEE International Conference on Acoustics, Speech and Signal Processing (ICASSP)}. IEEE, 2025, pp. 1--5.

\bibitem{gong2022contrastive}
Yuan Gong, Andrew Rouditchenko, Alexander~H Liu, David Harwath, Leonid Karlinsky, Hilde Kuehne, and James Glass,
\newblock ``Contrastive audio-visual masked autoencoder,''
\newblock {\em arXiv preprint arXiv:2210.07839}, 2022.

\bibitem{kim2019audiocaps}
Chris~Dongjoo Kim, Byeongchang Kim, Hyunmin Lee, and Gunhee Kim,
\newblock ``Audiocaps: Generating captions for audios in the wild,''
\newblock in {\em Proceedings of the 2019 Conference of the North American Chapter of the Association for Computational Linguistics: Human Language Technologies, Volume 1 (Long and Short Papers)}, 2019, pp. 119--132.

\bibitem{papineni2002bleu}
Kishore Papineni, Salim Roukos, Todd Ward, and Wei-Jing Zhu,
\newblock ``Bleu: a method for automatic evaluation of machine translation,''
\newblock in {\em Proceedings of the 40th annual meeting of the Association for Computational Linguistics}, 2002, pp. 311--318.

\bibitem{lin2004rouge}
Chin-Yew Lin,
\newblock ``Rouge: A package for automatic evaluation of summaries,''
\newblock in {\em Text summarization branches out}, 2004, pp. 74--81.

\bibitem{banerjee2005meteor}
Satanjeev Banerjee and Alon Lavie,
\newblock ``Meteor: An automatic metric for mt evaluation with improved correlation with human judgments,''
\newblock in {\em Proceedings of the acl workshop on intrinsic and extrinsic evaluation measures for machine translation and/or summarization}, 2005, pp. 65--72.

\bibitem{vedantam2015cider}
Ramakrishna Vedantam, C~Lawrence~Zitnick, and Devi Parikh,
\newblock ``Cider: Consensus-based image description evaluation,''
\newblock in {\em Proceedings of the IEEE conference on computer vision and pattern recognition}, 2015, pp. 4566--4575.

\bibitem{anderson2016spice}
Peter Anderson, Basura Fernando, Mark Johnson, and Stephen Gould,
\newblock ``Spice: Semantic propositional image caption evaluation,''
\newblock in {\em European conference on computer vision}. Springer, 2016, pp. 382--398.

\bibitem{liu2017improved}
Siqi Liu, Zhenhai Zhu, Ning Ye, Sergio Guadarrama, and Kevin Murphy,
\newblock ``Improved image captioning via policy gradient optimization of spider,''
\newblock in {\em Proceedings of the IEEE international conference on computer vision}, 2017, pp. 873--881.

\bibitem{wang2023internvid}
Yi~Wang, Yinan He, Yizhuo Li, Kunchang Li, Jiashuo Yu, Xin Ma, Xinhao Li, Guo Chen, Xinyuan Chen, Yaohui Wang, et~al.,
\newblock ``Internvid: A large-scale video-text dataset for multimodal understanding and generation,''
\newblock {\em arXiv preprint arXiv:2307.06942}, 2023.

\bibitem{ishikawa2025language}
Yuchi Ishikawa, Shota Nakada, Hokuto Munakata, Kazuhiro Saito, Tatsuya Komatsu, and Yoshimitsu Aoki,
\newblock ``Language-guided contrastive audio-visual masked autoencoder with automatically generated audio-visual-text triplets from videos,''
\newblock {\em arXiv preprint arXiv:2507.11967}, 2025.

\bibitem{li2023blip}
Junnan Li, Dongxu Li, Silvio Savarese, and Steven Hoi,
\newblock ``Blip-2: Bootstrapping language-image pre-training with frozen image encoders and large language models,''
\newblock in {\em International conference on machine learning}. PMLR, 2023, pp. 19730--19742.

\bibitem{agarwal2025gpt}
Sandhini Agarwal, Lama Ahmad, Jason Ai, Sam Altman, Andy Applebaum, Edwin Arbus, Rahul~K Arora, Yu~Bai, Bowen Baker, Haiming Bao, et~al.,
\newblock ``gpt-oss-120b \& gpt-oss-20b model card,''
\newblock {\em arXiv preprint arXiv:2508.10925}, 2025.

\end{thebibliography}

\end{document}